# Entropic measure and hypergraph states

Ri Qu, Yi-ping Ma, Yan-ru Bao, Juan Wang, and Zong-shang Li

*School of Computer Science and Technology, Tianjin University, Tianjin, 300072, China and*

*Tianjin Key Laboratory of Cognitive Computing and Application, Tianjin, 300072, China*

We investigate some properties of the entanglement of hypergraph states in purely hypergraph theoretical terms. We first introduce an approach for computing local entropic measure on qubit $t$ of a hypergraph state by using the Hamming weight of the so-called $t$-adjacent subhypergraph. Then we quantify and characterize the entanglement of hypergraph states in terms of local entropic measures obtained by using the above approach. Our results show that a class of $n$-qubit hypergraph states can not be converted into any graph state under local unitary transformations.

## I. INTRODUCTION

The understanding of the subtle properties of multipartite entangled states [1] is at the very heart of quantum information theory [2]. But the ultimate goal to cope with the properties of arbitrary multipartite states is far from being reached. Therefore, several special classes of entangled states have been introduced and identified to be useful for certain tasks. It is well known that *graph states* [3, 4] are an example of these classes. Any graph state can be constructed on the basis of a (simple and undirected) graph. Although graph states can describe a large family of entangled states including *cluster states* [5], *GHZ states*, *stabilizer states* [6], etc., it is clear that they cannot represent all entangled states. To go beyond graph states and still keep the appealing connection to graphs, Ref. [7] introduces an axiomatic framework for mapping graphs to quantum states of a suitable physical system, and extends this framework to directed graphs and weighted graphs. Several classes of multipartite entangled states, such as *qudit graph states* [8], *Gaussian cluster states* [9], *projected entangled pair states* [10], and *quantum random networks* [11], emerge from the axiomatic framework. In [12], we generalize the above axiomatic framework to encoding hypergraphs into so-called quantum hypergraph states.

It has been known that hypergraph states include graph states [12]. One may ask whether hypergraph states are equivalent to graph states under local unitary transformations, that is, whether hypergraph states can describe more quantum states than graph states under local unitary transformations. Ref. [13] has shown that one class of three-qubit hypergraph states can not be converted into any graph state under local unitaries. The main aim of this work is to answer the above question for $n$ qubits. For this, we will address the issue of using the *entropic measure* [14] to quantify and characterize the entanglement of hypergraph states in purely hypergraph theoretical terms. Several literatures have shown that there are several approaches for studying the properties of the entanglement of hypergraph states in the hypergraph theoretical terms. For graph states, Ref. [3] presents various upper and lower bounds to the *Schmidt measure* [15] in graph theoretical terms. For hypergraph states, similar work is done in [12]. Moreover, Ref. [12] qualitatively studies the entanglement structure of hypergraph states in purely hypergraph theoretical terms. In this paper, we will present an approach for computing local entropic measure on qubit $t$ of a hypergraph state by using the Hamming weight of the so-called $t$-adjacent subhypergraph. Then we will investigate some properties of the entanglement by using local entropic measures. Furthermore, we will show a class of *n*-qubit hypergraph states is not equivalent to any graph state under local unitary transformations.

This paper is organized as follows. In Sec. II, we recall notations of hypergraphs, hypergraph states, etc. In Sec. III, we define the Hamming weight of a hypergraph. Then we show how to calculate it by using the hypergraph theoretical terms. In Sec. IV, we present an approach for computing local entropic measures of a hypergraph state by means of the Hamming weights of some special subhypergraphs. Furthermore, we investigate some properties of the entanglement of hypergraph states by means of local entropic measures. We also indicate that a family of hypergraph states can not be converted into any graph state under local unitary transformations. Section V contains our conclusions.

## II. PRELIMINARIES

Formally, a *hypergraph* is a pair $(V, E)$, where $V$ is the set of *vertices*, $E \subseteq \wp(V)$ is the set of *hyperedges* and $\wp(S)$ denotes the power set of the set *S*. The *empty hypergraph* is defined as $(V, \varnothing)$. If a hypergraph only contains the *empty hyperedge* $\varnothing$ or one-vertex hyperedges (called *loops*), it is *trivial*. The *rank* of a hypergraph *g*, denoted by $ran(g)$, is the maximum cardinality of a hyperedge in *g*. A hypergraph $(V', E')$ is called a *subhypergraph* of $(V, E)$ if $V' \subseteq V$ and $E' \subseteq E$. For a vertex $t \in V$ we define the *t-adjacent subhypergraph* $g_t$ of $g = (V, E)$ as $g_t = (V_t, E_t)$ where $V_t = V - \{t\}$ and $E_t = \{e - \{t\} | t \in e \wedge e \in E\}$. Moreover, a hypergraph can be depicted by the visual form as shown in Fig. 1. Each vertex is represented as a dot while each hyperedge is represented as a closed curve which encloses the dots corresponding to vertices incident with the hyperedge.

Let $Z_k$ be the $2^k \times 2^k$ diagonal matrix which satisfies

$$(Z_k)_{jj} = \begin{cases} -1 & j = 2^k \\ 1 & others \end{cases} \quad (1)$$

where *k* is a nonnegative integer. Suppose that $V = [n] \equiv \{1, 2, ..., n\}$ and $e \subseteq V$. Then the *n*-qubit *hypergraph gate* $Z_e$ is defined as $Z_{|e|} \otimes I^{\otimes n-|e|}$ which means that $Z_{|e|}$ acts on the qubits in *e* while the identity *I* acts on the rest. An *n*-qubit *hypergraph state* $|g\rangle$ can be constructed by $g = (V, E)$ as follows. Each vertex labels a qubit (associated with a Hilbert space $\mathbb{C}^2$) initialized in $|\phi\rangle = |+\rangle \equiv \frac{1}{\sqrt{2}}(|0\rangle + |1\rangle)$. The state $|g\rangle$ is obtained from the initial state $|+\rangle^{\otimes n}$ by applying the hyperedge operator $Z_e$ for each hyperedge $e \in E$, that is,

$$|g\rangle = \prod_{e \in E} Z_e |+\rangle^{\otimes n}. \tag{2}$$

Thus hypergraph states of $n$ qubits are corresponding to $(\mathbb{C}^2, |+\rangle, \{Z_k \mid 0 \leq k \leq n\})$ by the axiomatic approach while graph states are related with $(\mathbb{C}^2, |+\rangle, Z_2)$ [7, 12].

It is known that real equally weighted states [16] are equivalent to hypergraph states [12]. In fact, let $V = [n]$ and define a mapping $c$ on $\wp(V)$ as

$$\forall e \subseteq V, c(e) = \begin{cases} 1 & e = \Phi \\ \prod_{k \in e} x_k & e \neq \Phi \end{cases}. \tag{3}$$

Then we can construct a *1-1* mapping $u$ between hypergraphs and Boolean functions which satisfies $\forall g = (V, E)$,

$$u(g)(x_1, x_2, \ldots, x_n) = \bigoplus_{e \in E} c(e). \tag{4}$$

where $\oplus$ denotes the addition operator over $\mathbb{Z}_2$. Thus we have

$$|g\rangle = \prod_{e \in E} Z_e |+\rangle^{\otimes n} = \frac{1}{\sqrt{2^n}} \sum_{x=0}^{2^n-1} (-1)^{\bigoplus_{e \in E} c(e)} |x\rangle \equiv |\psi_{u(g)}\rangle \tag{5}$$

where $|\psi_{u(g)}\rangle$ is just the real equally weighted state associate with the Boolean function $u(g)$.

Let $g$ and $g'$ be two hypergraphs with $n$ vertices. We say that they are *LU equivalent* if there exist local unitaries $\{U_l\}_{l=1,2,\ldots,n}$ such that

$$|\phi\rangle = U_1 \otimes U_2 \otimes \ldots \otimes U_n |\varphi\rangle, \tag{6}$$

i.e., $|g\rangle$ and $|g'\rangle$ are equivalent under local unitary operations.

### III. HAMMING WEIGHT

In this section we discuss how to obtain the Hamming weight of a Boolean function by means of hypergraph theory. It is known that the Hamming weight of a Boolean function $f$ is defined as $|f^{-1}(1)|$ where $|S|$ denotes the cardinality of the set $S$. By (4), we also can define the Hamming weight of a hypergraph $g$ with $n$ vertices as

$$hw(g) \equiv |f^{-1}(1)| \tag{7}$$

where $f(x_1, x_2, ..., x_n) = u(g)(x_1, x_2, ..., x_n)$. We give the following proposition to calculate the Hamming weight of $g$.

*Propostion 1.* Let $g = ([n], E)$ be a hypergraph with $n$ vertices. Then

(i) If $E = \emptyset$, then $hw(g) = 0$.

(ii) If $E = \{e_1, e_2, ..., e_m\}$, then

$$hw(g) = \sum_{i=1}^{m} 2^{n-|e_i|} - 2\sum_{1 \leq i < j \leq m} 2^{n-|e_i \cup e_j|} + 4\sum_{1 \leq i < j < k \leq m} 2^{n-|e_i \cup e_j \cup e_k|} + ... + (-2)^{m-1} 2^{n-|e_1 \cup e_2 \cup ... \cup e_m|}. \quad (8)$$

*Proof.* (i) Form (4), we can obtain that $u(g)(x_1, x_2, ..., x_n) = 0$ for any $x_1, x_2, ..., x_n$. Thus $hw(g) = 0$ by (7). (ii) When $m = 1$, it is easily seen that (8) is true by (4) and (7). Assume that (8) is true when $m = t$. Now we will prove that (8) is also true when $m = t+1$. Let $E = \{e_1, e_2, ..., e_{t+1}\}$ and $g' = ([n], E - \{e_{t+1}\})$. We denote $hw(g) - hw(g')$ by $\Delta$. It is clear that

$$\Delta = 2^{n-|e_{t+1}|} - 2\sum_{i=1}^{t} 2^{n-|e_i \cup e_{t+1}|} + 4\sum_{1 \leq i < j \leq t} 2^{n-|e_i \cup e_j \cup e_{t+1}|} + ... + (-2)^t 2^{n-|e_1 \cup e_2 \cup ... \cup e_{t+1}|}. \quad (9)$$

Since $g'$ has $t$ hyperedges, we can obtain

$$hw(g') = \sum_{i=1}^{t} 2^{n-|e_i|} - 2\sum_{1 \leq i < j \leq t} 2^{n-|e_i \cup e_j|} + 4\sum_{1 \leq i < j < k \leq t} 2^{n-|e_i \cup e_j \cup e_k|} + ... + (-2)^{t-1} 2^{n-|e_1 \cup e_2 \cup ... \cup e_t|}. \quad (10)$$

By (9) and (10), we can obtain the equation (8) for $m = t+1$. ∎

According to (8), it is easy to obtain the Hamming weight of a hypergraph. For instance, $hw(g_4)$ of the hypergraph $g_4$ in Fig. 1(d) is equal to 6 by (8). According to the above proposition, the Hamming weight of a hypergraph has some properties as follows. We will first discuss how to identify whether the Hamming weight of a hypergraph is odd or not.

*Corollary 2.* Let $g = ([n], E)$ be a hypergraph. $hw(g)$ is odd if and only if $[n] \in E$.

*Proof.* (i) "if". Suppose that $E = \{e_1, e_2, ..., e_m\}$. Without loss of generality, let us denote the hyperedge $[n]$ by $e_m$. If $m = 1$, then $E = \{[n]\}$. Thus it clear that $hw(g) = 1$. If $m \geq 2$, we denote the hypergraph $([n], E - \{[n]\})$ by $g'$. According to the above proposition, we can obtain that

$$hw(g') = \sum_{i=1}^{m-1} 2^{n-|e_i|} - 2\sum_{1 \leq i < j \leq m-1} 2^{n-|e_i \cup e_j|}$$

$$+4\sum_{1\leq i<j<k\leq m-1}2^{n-|e_i\cup e_j\cup e_k|}+\ldots+(-2)^{m-2}\,2^{n-|e_1\cup e_2\cup\ldots\cup e_{m-1}|}. \tag{11}$$

Clearly, $hw(g)$ is even. By (8), it is known that

$$hw(g)=hw(g')+\Delta \tag{12}$$

where

$$\Delta=2^{n-\|[n]\|}-2\sum_{1\leq i\leq m-1}2^{n-|e_i\cup[n]|}+4\sum_{1\leq i<j\leq m-1}2^{n-|e_i\cup e_j\cup[n]|}+\ldots+(-2)^{m-1}2^{n-|e_1\cup e_2\cup\ldots\cup e_{m-1}\cup[n]|}. \tag{13}$$

By simple calculation, we can obtain

$$\Delta=C_{m-1}^0-2C_{m-1}^1+4C_{m-1}^2+\ldots+(-2)^{m-1}C_{m-1}^{m-1}=(-1)^{m-1}. \tag{14}$$

Thus $hw(g)$ is odd since $hw(g')$ is even and $\Delta$ is odd. (ii) "only if". Assume that $[n]\notin E$. Clearly, we can obtain that $hw(g)$ would be even by (8). ∎

According to the proof of the above corollary, we can easily obtain the following corollary.

*Corollary 3.* Let $g=([n],E)$ be a hypergraph. If $[n]\in E$, then

$$hw(g)=hw(g')+(-1)^{m-1}. \tag{15}$$

where $g'=([n],E-\{[n]\})$ and $m=|E|$.

We say that a hypergraph is odd (even) if its Hamming weight is odd (even). By the corollary 2, it is known that the sufficiency and necessary condition of an *n*-vertex odd hypergraph $g$ is $\operatorname{ran}(g)=n$. In the following, we will investigate the properties of the hamming weight of the hypergraph $g$ for $\operatorname{ran}(g)<n$.

*Proposition 4.* Let $g$ be a hypergraph with $n$ vertices. Then

(i) $\operatorname{ran}(g)=0$ if and only if $hw(g)$ equals to either 0 or $2^n$.

(ii) if $\operatorname{ran}(g)=1$, then $hw(g)=2^{n-1}$.

*Proof.* It is clear that (i) is true. Now we prove (ii). Let $g=([n],E)$ and $m=|E|$. For $\operatorname{ran}(g)=1$, two cases should be considered as follows. (*) The empty hyperedge $\varnothing\notin E$. It is clear that $m\geq 1$ and each hyperedge in $g$ is a loop. According to the proposition 1, we can obtain

$$hw(g)=\sum_{i=1}^m 2^{n-1}-2\sum_{1\leq i<j\leq m}2^{n-2}+4\sum_{1\leq i<j<k\leq m}2^{n-3}+\ldots+(-2)^{m-1}2^{n-m}$$

$$=C_m^1 2^{n-1}-C_m^2 2^{n-1}+C_m^3 2^{n-1}+\ldots+(-1)^{m-1}C_m^m 2^{n-1}$$

$$= -2^{n-1}\left[-C_m^1 + C_m^2 - C_m^3 + \ldots + (-1)^m C_m^m\right]$$

$$= -2^{n-1}\left[C_m^0 - C_m^1 + C_m^2 - C_m^3 + \ldots + (-1)^m C_m^m - 1\right]$$

$$= -2^{n-1}\left[(1-1)^m - 1\right] = 2^{n-1}. \tag{16}$$

(**) The empty hyperedge $\varnothing \in E$. It is clear that $m \geq 2$ and all hyperedges except the empty hyperedge $\varnothing$ are loops. Thus it is clear that

$$hw(g) = 2^n - hw(g') \tag{17}$$

where $g' = ([n], E - \{\varnothing\})$. According to (i), it is known that $hw(g') = 2^{n-1}$ since $\varnothing$ is not included in $g'$. By (17), we can obtain $hw(g) = 2^{n-1}$. ∎

Note that the converse proposition of (ii) in the above proposition is not true. In fact, it is clear that the Hamming weight of a *3*-vertex complete graph $g = (\{1,2,3\}, E)$ with $E = \{\{1,2\},\{1,3\},\{2,3\}\}$ is equal to *4* while its rank is *2*.

## IV. LOCAL ENTROPIC MEASURES

In this paper, the local entropic measure $E_2^t(|\phi\rangle)$ on qubit $t$ of an $n$-qubit pure state $|\phi\rangle$ is given by the determinant of the reduced density matrix $\rho_t \equiv \text{Tr}_{\text{all but } t}(|\phi\rangle\langle\phi|)$, that is, $E_2^t(|\phi\rangle) \equiv \det(\rho_t)$ [17]. It is known that the local entropic measures are an entanglement monotone for a single copy case and invariant under local unitary operations. Moreover, it is clear that $0 \leq E_2^t(|\phi\rangle) \leq \frac{1}{4}$.

Let $g = ([n], E)$ be a hypergraph. By (5), the reduced state on qubit $t$ of the corresponding hypergraph state $|g\rangle$ can be written into

$$\rho_t = \text{Tr}_{\text{all but } t}(|g\rangle\langle g|) = \begin{bmatrix} \frac{1}{2} & a \\ a & \frac{1}{2} \end{bmatrix} \tag{18}$$

where $a = \frac{1}{2^n} \sum_{x_1,\ldots,x_{t-1},x_{t+1},\ldots,x_n=0}^{1} (-1)^{u(g)(x_1,\ldots,x_{t-1},0,x_{t+1},\ldots,x_n) \oplus u(g)(x_1,\ldots,x_{t-1},1,x_{t+1},\ldots,x_n)}$. It is known that there are two ($n$-$1$)-valuable Boolean functions $p$ and $q$ such that

$$u(g)(x_1, x_2, \ldots, x_n) = x_t p(x_1, \ldots, x_{t-1}, x_{t+1}, \ldots, x_n) \oplus q(x_1, \ldots, x_{t-1}, x_{t+1}, \ldots, x_n). \tag{19}$$

Then we can obtain

$$a = \frac{1}{2^n} \sum_{x_1,\ldots,x_{t-1},x_{t+1}\ldots,x_n=0}^{1} (-1)^{p(x_1,\ldots,x_{t-1},x_{t+1}\ldots,x_n)}. \tag{20}$$

By the definition of the *t*-adjacent subhypergraph and (19), it is clear that

$$p(x_1,\ldots,x_{t-1},x_{t+1}\ldots,x_n) = u(g_t). \tag{21}$$

From (20) and (21), we can obtain

$$a = \frac{1}{2^n}\left[2^{n-1} - 2hw(g_t)\right]. \tag{22}$$

According to the definition of the local entropic measure, we can get

$$E_2^t(|g\rangle) = \frac{1}{4} - a^2. \tag{23}$$

Thus it is import for calculating $E_2^t(|g\rangle)$ to obtain the Hamming weight of the *t*-adjacent subhypergraph $g_t$ of $g$. As shown in Fig. 1, the hypergraph $g_4$ in (d) is just the *4*-adjacent subhypergaph of $g$ in (a). Since it is shown that $hw(g_4) = 6$ in Sec. III, we can obtain that $E_2^4(|g\rangle) = \frac{3}{16}$ by (22) and (23). Next, we will discuss some properties of local entropic measures in the hypergraph theoretical terms.

*Propositon 5.* Let $g = ([n], E)$ be a hypergraph and $t \in [n]$. Then

(i) $\operatorname{ran}(g_t) = 0$ if and only if $E_2^t(|g\rangle) = 0$.

(ii) if $\operatorname{ran}(g_t) = 1$, then $E_2^t(|g\rangle) = \frac{1}{4}$, that is, qubit *t* is maximally entangled with the other *n-1* qubits.

(iii) if $\operatorname{ran}(g_t) = n-1$, then $0 < E_2^t(|g\rangle) < \frac{1}{4}$.

*Proof.* (i) (*)"if". Since $E_2^t(|g\rangle) = 0$, the absolute of $a$ in (23) has to be $\frac{1}{2}$. Then, by (22), $hw(g_t)$ is either 0 or $2^{n-1}$. This implies that $\operatorname{ran}(g_t) = 0$ according to the proposition 4.

(**)"only if". Since $\operatorname{ran}(g_t) = 0$, we can obtain that $hw(g_t)$ is either 0 or $2^{n-1}$ according to the proposition 4. Then the reduced density operator $\rho_t$ can be written into one of

$$\begin{bmatrix} \frac{1}{2} & \frac{1}{2} \\ \frac{1}{2} & \frac{1}{2} \end{bmatrix} \text{ and } \begin{bmatrix} \frac{1}{2} & -\frac{1}{2} \\ -\frac{1}{2} & \frac{1}{2} \end{bmatrix}. \tag{24}$$

Thus $E_2^t(|g\rangle)=0$ by (23). (ii) Since $\text{ran}(g_t)=1$, we can obtain that $hw(g_t)=2^{n-2}$ according to the proposition 4. Then, by (22) the reduced density operator $\rho_t$ can be written into

$$\rho_t = \begin{bmatrix} \frac{1}{2} & 0 \\ 0 & \frac{1}{2} \end{bmatrix}.$$

Thus $E_2^t(|g\rangle)=\frac{1}{4}$ by (23). (iii) Since $\text{ran}(g_t)=n-1$, we can obtain that $hw(g_t)$ is odd by the corollary 2. Thus $0<E_2^t(|g\rangle)<\frac{1}{4}$. ∎

Note that the converse proposition of (ii) in the above proposition is not true. For $g'$ shown in Fig. 1(b), it is clear that $E_2^4(|g'\rangle)=\frac{1}{4}$ while $\text{ran}(g_4')=2>1$. In fact, $hw(g_4')=4$ since $g_4'$ is a 3-vertex complete graph, which is shown in Sec. III. By (22) and (23) we can obtain

$$E_2^4(|g'\rangle)=\frac{1}{4}.$$

We say that an $n$-qubit pure state $|\phi\rangle$ is locally maximally entangled [18] if all of its local entropic measures are the maximum, that is, for any $t\in[n]$ it holds that $E_2^t(|\phi\rangle)=\frac{1}{4}$. In Fig. 1(c), the state $|g''\rangle$ is just of graph state and it is locally maximally entangled by the above proposition. It is easily seen that all graph states are locally maximally entangled. One may ask whether every locally maximally entangled hypergraph belongs to graph states. Our answer is "no". In fact, the hypergraph $(\{1,2,3,4\},E)$ with $E=\{\{i,j,k\}\,|\,1\leq i<j<k\leq 4\}$ is locally maximally entangled but not of graph states. Moreover, the following corollary can be easily obtained according to the above proposition.

*Corollary 6.* Let $g$ be a hypergraph with $n$ vertices. Then

(i) If $\text{ran}(g)\in\{0,1\}$, then $E_2^t(|g\rangle)=0$ for any vertex $t$.

(ii) If $\text{ran}(g)=2$, then $E_2^t(|g\rangle)\in\left\{0,\frac{1}{4}\right\}$ for any vertex $t$.

(iii) If $\text{ran}(g)=n$, then $0<E_2^t(|g\rangle)<\frac{1}{4}$ for any vertex $t$.

*Proposition 7.* Let $g=([n],E)$ and $g'=([n],E')$ be two hypergraphs with $n$ vertices. If $\text{ran}(g)=n$ and $\text{ran}(g')\leq n-1$, then $E_2^t(|g\rangle)\neq E_2^t(|g'\rangle)$ for any vertex $t$.

*Proof.* Since $\operatorname{ran}(g) = n$ and $\operatorname{ran}(g') \leq n-1$, we can obtain that $[n] \in E$ and $[n] \notin E'$. According to the corollary 2, it is known that $hw(g_t)$ and $hw(g_t')$ are respectively odd and even for any vertex *t*. By (22) and (23), it is clear that $E_2^t(|g\rangle) \neq E_2^t(|g'\rangle)$. ∎

By the above proposition and the properties of local entropic measure, we can obtain main conclusion of this work as follows.

*Proposition 8.* Let $g = ([n], E)$. If $[n] \in E$, the hypergraph *g* is not LU equivalent to any hypergraph which does not include the hyperedge $[n]$.

Suppose that $n \geq 3$. It is clear that no graph with *n* vertices includes the hyperedge $[n]$. Thus no graph state of *n* qubits is LU equivalent to any *n*-qubit hypergraph state whose corresponding hypergraph $g = ([n], E)$ satisfies $[n] \in E$.

## V. CONCLUSIONS

This work uses the local entropic measures to quantify and characterize the entanglement of hypergraph states of *n* qubits. For this, we introduce an approach for computing the local entropic measures in purely hypergraph theoretical terms. At first, we define the Hamming weight of a hypergraph. Then we give a method to compute the Hamming weight of the hypergraph by using hypergraph theory. And we prove that one can use the Hamming weight of the *t*-adjacent subhypergraph to calculate the local entropic measure on qubit *t* of a hypergraph state. Our research shows that the *n*-vertex hypergraphs including the hyperedge $[n]$ are not LU equivalent to any other hypergraph. Since (simple and undirected) graphs with *n* $(n \geq 3)$ vertices do not include the hyperedge $[n]$, hypergraph states of *n* qubits are not equivalent to graph states under local unitaries. This implies that hypergraph states can represent more states than graph stares under local unitary transformations.

**ACKNOWLEDGMENTS**

This work is supported by the Chinese National Program on Key Basic Research Project (973 Program, Grant No. 2013CB329304) and the Natural Science Foundation of China (Grant Nos. 61170178, 61272254, 61105072 and 61272265). This work is completed during our academic visiting at Department of Computing, Open University, UK.

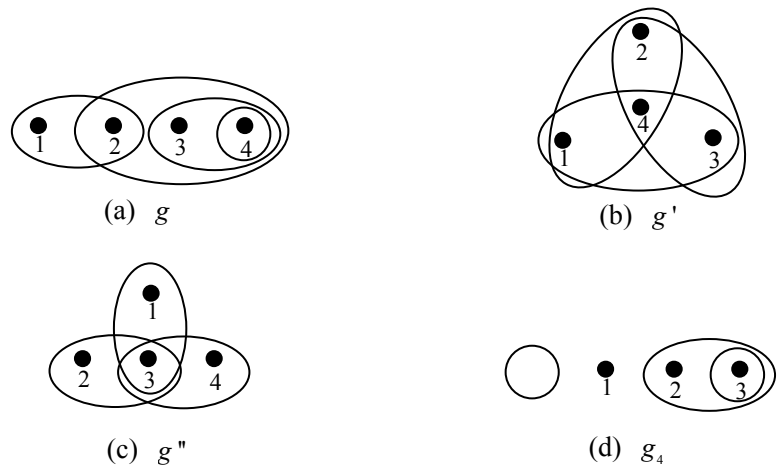

Figure 1. Examples of hypergraphs. The hypergraphs (a)-(c) have the same vertex set $\{1,2,3,4\}$. The hypergraph $g$ in (a) has *4* hyperedges: $\{4\},\{1,2\},\{3,4\}$ and $\{2,3,4\}$. In (b), the hypergraph $g'$ also has *3* hyperedges: $\{1,2,4\}$, $\{1,3,4\}$ and $\{2,3,4\}$. Three hyperedges, i.e., $\{1,3\}$, $\{2,3\}$ and $\{3,4\}$, constitute the hyperedge set of $g''$ in (c). The *3*-vertex hypergraph $g_4$ in (d) has *3* hyperedges: $\varnothing$, $\{3\}$ and $\{2,3\}$. The hypergraph $g_4$ is just the *4*-adjacent subhypergaph of $g$.